\documentclass[conference]{IEEEtran}
\pdfoutput=1
\IEEEoverridecommandlockouts

\usepackage{makeidx}  
\usepackage{color}
\usepackage{graphicx}
\usepackage{amsfonts}
\usepackage{url}
\usepackage{amsmath}
\usepackage{listings}
\usepackage{amsthm}
\usepackage{xspace}
\usepackage{hyperref}
\usepackage{subfigure}

\usepackage[noend]{algpseudocode} 
\algrenewcommand\algorithmicindent{0.7em}

\algdef{SE}[PROCEDURE]{ProcedureLi}{EndProcedureLi}%
[3]{\algorithmicprocedure\ \textproc{#1}\ifthenelse{\equal{#2}{}}{}{(#2)}%
,\quad\quad \textbf{returns}: {#3}}%
{\algorithmicend\ \algorithmicprocedure}%

\usepackage{booktabs}
\usepackage{rotating}

\newcommand{\true}{\mathbf{true}}
\newcommand{\false}{\mathbf{false}}
\DeclareMathOperator{\scope}{\mathbin{.}}
\newcommand{\depqbf}{\textsf{DepQBF}\xspace}
\newcommand{\bloqqer}{\textsf{Bloqqer}\xspace}
\newcommand{\demiurge}{\textsf{Demiurge}\xspace}

\newcommand{\lingeling}{\textsf{Lingeling}\xspace}
\newcommand{\picosat}{\textsf{PicoSat}\xspace}
\newcommand{\mathsat}{\textsf{MathSAT}\xspace}

\newcommand{\qbfcert}{\textsf{QBFCert}\xspace}
\newcommand{\cudd}{\textsf{CuDD}\xspace}
\newcommand{\Abc}{\textsf{ABC}\xspace}
\newcommand{\aiger}{\textsf{AIGER}\xspace}

\newcommand{\propsatmodel}{\textsc{PSat}}
\newcommand{\propsatcore}{\textsc{PCore}}
\newcommand{\interpol}{\textsc{Int}}
\newcommand{\subby}{\kern-0.2em\leftarrow\kern-0.2em}

\newcommand{\qbfsatmodel}{\textsc{QSat}}
\newcommand{\qbfsatcore}{\textsc{QCore}}

\begin{document}
%
\title{SAT-Based Methods for Circuit Synthesis
\thanks{This work was
supported in part by the Austrian Science Fund (FWF) through the
national research network RiSE (S11406-N23, S11409-N23) and the project QUAINT
(I774-N23), as well as by the European Commission through project
STANCE (317753).}}

\author{\IEEEauthorblockN{Roderick Bloem$^1$,
                          Uwe Egly$^2$,
                          Patrick Klampfl$^1$, 
                          Robert K\"{o}nighofer$^1$, and
                          Florian Lonsing$^2$
                          }
\vspace{0.2cm}
\IEEEauthorblockA{
$^1$Institute for Applied Information Processing and Communications,
Graz University of Technology, Austria}
\IEEEauthorblockA{
$^2$Knowledge-Based Systems Group, Institute of Information Systems,
Vienna University of Technology, Austria}
}

\maketitle

\begin{abstract}
Reactive synthesis supports designers by automatically constructing correct 
hardware from declarative specifications.  Synthesis algorithms usually compute 
a strategy, and then construct a circuit that implements it. In this work, we 
study SAT- and QBF-based methods for the second step, i.e., computing circuits 
from strategies. This includes methods based on QBF-certification, 
interpolation, and computational learning. We present optimizations, efficient 
implementations, and experimental results for synthesis from safety 
specifications, where we outperform BDDs both regarding execution time and 
circuit size.
This is an extended version of \cite{fmcad14}, with an
additional appendix.
\end{abstract}

\IEEEpeerreviewmaketitle

\section{Introduction} \label{sec:intro}
Synthesis is an ambitious design approach:  Instead of checking whether an 
already constructed system satisfies its specification, a correct implementation 
is derived \emph{automatically} from the specification~\cite{BloemGJPPW07}.  
Synthesis is also used in rapid prototyping, automatic 
repair~\cite{JobstmannSGB12}, and program sketching~\cite{Solar-Lezama09}.

Existing work often focuses on finding strategies to satisfy the specification, 
or only on deciding realizability. However, computing circuits from strategies 
is computationally demanding as well.  System quality (e.g., circuit size and 
depth) imposes additional challenges. Synthesized strategies usually allow for 
much implementation freedom, which needs to be exploited cleverly. Algorithms 
must also be symbolic (operate on formulas rather than enumerating states) to 
achieve scalability.  These symbolic algorithms are usually implemented with 
BDDs because they offer existential \emph{and} universal quantification. 
Recently, SAT-based synthesis algorithms have been 
proposed~\cite{MorgensternGS13,BloemKS14} as alternative to BDDs and their 
scalability issues. However, these works do not address circuit extraction.

We thus present and compare several SAT- and QBF-based circuit synthesis 
algorithms.  The basic algorithms are not new, but we present novel 
optimizations, combinations, efficient implementations for safety synthesis 
problems, and extensive experiments.  This includes methods based on 
QBF-certification, computational learning (including the first application of 
incremental QBF solving in synthesis), and interpolation.  We achieve the best 
results by combining ideas from interpolation~\cite{JiangLH09} with 
learning~\cite{EhlersKH12}, thereby outperforming BDDs both regarding 
computation time and circuit size.

\textbf{Related work.} It is argued~\cite{EhlersKH12} that many circuit 
synthesis methods are still outperformed by the simple BDD-based cofactor 
approach~\cite{BloemGJPPW07}. The same work~\cite{EhlersKH12} also proposes 
learning-based techniques, which are implemented with BDDs.  This yields smaller 
circuits, but is slower.  We show how learning can be efficiently realized with 
SAT- and QBF-solvers, and that this realization can outperform the cofactor 
approach both regarding circuit size and computation time.  SAT-based learning 
is also used in~\cite{BloemKS14}.  However, this work only addresses strategy 
computation and not circuit synthesis.  Jiang et al.~\cite{JiangLH09} propose 
interpolation for circuit extraction, and show how quantifier alternations can 
be avoided by temporarily treating outputs as inputs. We combine this idea with 
learning to compute interpolants, thereby achieving a speedup of several orders 
of magnitude.  QBF certification~\cite{NiemetzPLSB12} can derive circuits from a 
completeness proof of the strategy formula.  We show how this method can be 
applied efficiently for safety synthesis.

\section{Preliminaries} \label{sec:prel}

We assume familiarity with propositional logic, SAT- and QBF-solving 
(cf.~\cite{hos}) but repeat the most important concepts.

\textbf{Basic Notation.}
A \emph{literal} is a Boolean variable or its negation. A \emph{clause} 
(\emph{cube}) is a disjunction (conjunction) of literals, and a 
\emph{Conjunctive Normal Form (CNF)} formula is a conjunction of clauses.  We 
denote variables vectors with overlines, corresponding cubes in bold, and 
propositional formulas with capital letters. E.g., $\mathbf{x}$ is a cube over 
the variable vector $\overline{x}=(x_1,\ldots,x_n)$, and $F(\overline{x})$ is a 
formula over $\overline{x}$.  If the variables are irrelevant, we simply write 
$F$ instead of $F(\overline{x})$.

\textbf{Decision Procedures.}
A \emph{SAT-solver} checks if a CNF is satisfiable.  We write $(\textsf{sat}, 
\mathbf{x}) := \propsatmodel(F(\overline{x}))$ for a SAT-solver call, where 
$\textsf{sat}$ is assigned $\true$ iff the CNF $F$ is satisfiable, and 
$\mathbf{x}$ is a satisfying assignment given as cube over $\overline{x}$.  Let 
$\mathbf{x}$ be a cube.  We write $\mathbf{y} := \propsatcore(\mathbf{x}, F)$ to 
denote the extraction of an unsatisfiable core: Given that $\mathbf{x} \wedge F$ 
is unsatisfiable, $\mathbf{y}$ will be a sub-cube of $\mathbf{x}$ such that 
$\mathbf{y} \wedge F$ is still unsatisfiable.  Let 
$A(\overline{x},\overline{y})$ and $B(\overline{x},\overline{z})$ be two CNFs 
such that $A\wedge B$ is unsatisfiable, and $\overline{y}$ and $\overline{z}$ 
are disjoint.  An \emph{interpolant} is a formula $I(\overline{x})$ such that $A 
\Rightarrow I \Rightarrow \neg B$.  Interpolants can be computed from the 
unsatisfiability proof of $A\wedge B$~\cite{DSilvaKPW10}.  We denote this 
computation by $I := \interpol(A,B)$.
A \emph{Quantified Boolean Formula (QBF)} is a formula $Q_1\overline{x} \scope 
Q_2\overline{y} \ldots F(\overline{x},\overline{y}, \ldots)$, where $Q_i \in 
\{\forall, \exists\}$ and $F$ is a CNF.  The quantifiers have their expected 
semantics.  A \emph{QBF-solver} checks if a QBF is satisfiable. We write
$(\textsf{sat}, \mathbf{a}) :=
\qbfsatmodel(
\exists\overline{a} \scope
Q_1 \overline{x} \scope
Q_2 \overline{y} \ldots
F(\overline{a}, \overline{x},\overline{y}, \ldots))$
for QBF-solver calls.  The satisfying assignment $\mathbf{a}$ can only be 
extracted for variables that are quantified existentially on the outermost 
level. Finally, we write
$\mathbf{b} :=
\qbfsatcore(\mathbf{a},
\exists\overline{a} \scope
Q_1 \overline{x} \scope
Q_2 \overline{y} \ldots
F(\overline{a},
\overline{x},\overline{y}, \ldots))$
to denote the extraction of an unsatisfiable core.

\begin{figure}[tb]
 \centering
   \includegraphics[width=0.4\textwidth]{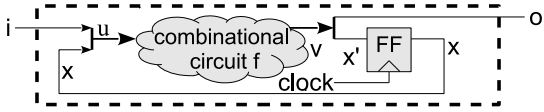}
   \caption{Implementation of a strategy. (FF = flip-flops).}
   \label{fig:impl}
\end{figure}

\textbf{Circuit Synthesis.}
A \emph{strategy} is a formula $S(\overline{x},\overline{i},\overline{o}, 
\overline{x}')$ such that $\forall \overline{x}, \overline{i} \scope \exists 
\overline{o}, \overline{x}' \scope S$, where 
$\overline{x},\overline{i},\overline{o}$ are state-, input-, and output-bits, 
respectively, and $\overline{x}'$ is the next-state copy of $\overline{x}$. 
Intuitively, for a given state $\mathbf{x}$ and input $\mathbf{i}$, $S$ defines 
allowed output-values $\mathbf{o}$ and next states $\mathbf{x}'$: $\mathbf{o}, 
\mathbf{x}'$ is \emph{allowed} iff $\mathbf{x} \wedge \mathbf{i} \wedge 
\mathbf{o} \wedge \mathbf{x}'$ satisfies $S$.  Let $\overline{u} = \overline{x} 
\cup \overline{i}$ and $\overline{v} = \overline{o} \cup \overline{x}'$.  An 
\emph{implementation} of $S(\overline{u},\overline{v})$ is a function 
$f:2^{|\overline{u}|} \rightarrow 2^{|\overline{v}|}$ such that $\forall 
\overline{u} \scope S(\overline{u},f(\overline{u}))$.  This function can be 
implemented in hardware as shown in Fig.~\ref{fig:impl}.

Strategies for safety specifications are particularly simple: given a 
\emph{winning region} $W(\overline{x})$ from which the specification can  be 
enforced, and a complete\footnote{I.e., $\forall \overline{x}, \overline{i}, 
\overline{o} \scope \exists \overline{x}'\scope 
T(\overline{x},\overline{i},\overline{o},\overline{x}')$. $T$ can always be made 
complete: if some input is not allowed by the original specification, $T$ can 
allow for arbitrary outputs; if some output is not allowed originally, $T$ can 
visit an unsafe state.} and deterministic\footnote{I.e., $\forall \overline{x}, 
\overline{i}, \overline{o}, \overline{x_1}', \overline{x_2}' \scope 
(T(\overline{x},\overline{i},\overline{o},\overline{x_1}') \wedge 
T(\overline{x},\overline{i},\overline{o},\overline{x_2}')) \Rightarrow 
(\overline{x_1}' = \overline{x_2}') $.} transition relation 
$T(\overline{x},\overline{i},\overline{o},\overline{x}')$ defining the state 
transitions, the strategy must pick values for $\overline{o}$ such that the next 
state is in $W$ again, i.e.,
$S = \bigl(\neg W(\overline{x})\bigr)
     \vee \bigl(
     T(\overline{x},\overline{i},\overline{o},\overline{x}') \wedge
     W(\overline{x}') \bigr)$. 
We only need to synthesize circuits for $\overline{o}$, and define 
$\overline{x}'$ using $T$.

\section{Circuit Synthesis Algorithms} \label{sec:alg}

\subsection{QBF-Certification} \label{sec:qbfcert}

An implementation can be computed as Skolem functions\footnote{Skolem functions 
define existentially quantified variables as a function over the universally 
quantified ones such that the QBF becomes $\true$.} for the signals 
$\overline{o}$ and $\overline{x}'$ in the QBF $\forall \overline{x}, 
\overline{i} \scope \exists \overline{o}, \overline{x}' \scope 
S(\overline{x},\overline{i},\overline{o}, \overline{x}')$.  
\qbfcert~\cite{NiemetzPLSB12} computes such functions using 
\depqbf~\cite{LonsingB10}.

\textbf{Optimizations for Safety Specifications.}   We need to find Skolem 
functions for $\overline{o}$ in
$\forall \overline{x}, \overline{i} \scope \exists \overline{o}, \overline{x}' 
\scope (\neg W) \vee (T \wedge W')$.
Yet, we achieve better results with \qbfcert by computing Herbrand 
functions\footnote{Herbrand functions define universally quantified variables as 
a function over the existentially quantified ones such that the QBF becomes 
$\false$.} in the unsatisfiable QBF 
$\exists \overline{x}, \overline{i} \scope
\forall \overline{o} \scope \exists \overline{x}' \scope
W \wedge T \wedge \neg W'$.
This works because $T$ is deterministic and complete. In our setting, $W$ is in 
CNF, so the conjunctions in the latter formulation are simpler to realize in 
CNF. Also, the clause resolution proofs required for unsatisfiable QBFs are 
usually less expensive than the cube resolution proofs for satisfiable ones. 
Still, the intermediate files produced by \qbfcert can grow large (hundreds of 
GB). One reason is that a straightforward CNF encoding of $\neg W'$ requires 
many auxiliary variables and clauses.  We could reduce the size of the files by 
up to a factor of 30 by learning a CNF representation of $\neg W'$ without 
introducing auxiliary variables using the following algorithm:
\begin{algorithmic}[1]
\ProcedureLi{NegLearn}
             {$W'$}
             {$\neg W'$}
  \State $N' := \true$
  \While{$\mathsf{sat}$, with $(\mathsf{sat},\mathbf{x}):=
  \propsatmodel(W' \wedge N')$}
    \State $N' := N' \wedge \neg \propsatcore(\mathbf{x}, \neg W')$
  \EndWhile
  \State \textbf{return} $N'$
\EndProcedure
\end{algorithmic}
As long as $N'$ is not yet $\neg W'$, i.e., $W' \wedge N'$ is still satisfiable, 
we refine $N'$ with a clause that excludes the cube $\mathbf{x}$ witnessing this 
insufficiency.  By taking the unsatisfiable core, the clause eliminates also 
other counterexamples.  Since clauses are only added, \textsc{NegLearn} is 
suitable for incremental SAT solving.

Using incremental SAT solving, we also simplify $W$ by removing literals and 
clauses as long as $W$ does not change semantically. This is applied to all 
following methods as well.

\subsection{QBF-Based Learning} \label{sec:qbflearn}

In~\cite{EhlersKH12}, several learning-based circuit synthesis algorithms are 
presented and implemented using BDDs.  Here, we discuss an efficient 
implementation of the CNF-learning algorithm using a QBF-solver.  Since 
QBF-solvers operate on CNFs, this algorithm is particularly suitable.  It can be 
defined as follows.
\begin{algorithmic}[1]
\Procedure{SyLearnQBF}{$S(\overline{x},\overline{i},\overline{o},
                          \overline{x}')$}
  \State $\overline{u} := \overline{x} \cup \overline{i}$,\;\; 
         $\overline{v}_a := \overline{v} := \overline{o} \cup \overline{x}'$
  \For{$v\in \overline{v}$}
    \State $\overline{v}_a := \overline{v}_a \setminus \{v\}$,\;
           $\overline{v}_e := \overline{v} \setminus \overline{v}_a$,\;
           $f_v := \true$,\;
           $R := v \wedge \neg S$
    \While{$\mathsf{sat}$, with
      $(\mathsf{sat},\mathbf{u})$:=$\qbfsatmodel(
      \exists \overline{u} \scope
      \forall \overline{v}_a \scope
      \exists \overline{v}_e \scope R)$}\label{alg:qbfl:check}
      \State $\mathbf{u}_2 := \qbfsatcore(\mathbf{u},
          \exists \overline{u} \scope
          \forall \overline{o_a} \scope
          \exists \overline{o_e}, \overline{x}' \scope
          \neg v \wedge \neg S)$ \label{alg:qbfl:core}
      \State $f_v := f_v \wedge \neg \mathbf{u}_2$,\;\;
             $R := R \wedge \neg \mathbf{u}_2$
    \EndWhile
    \State $\textsc{dumpCircuit}(v, f_v)$,\;\;
           $S := S \wedge (v\leftrightarrow f_v)$ \label{alg:qbfl:resub}
  \EndFor
\EndProcedure
\end{algorithmic}
\textsc{SyLearnQBF} builds up circuits in $f_v$ for one $ v\in \overline{v}$ 
after the other.  Initially, $f_v=\true$, i.e., the circuit always outputs 
$\true$.  While there exists an input $\mathbf{u}$ for which $v$ must be 
$\false$ (the QBF in line \ref{alg:qbfl:check} is satisfiable), $f_v$ is refined 
with a clause that maps $\mathbf{u}$ to $\false$. By taking the core in 
line~\ref{alg:qbfl:core}, other inputs are also mapped to $\false$ as long as 
$\false$ is allowed by $S$. The final solution $f_v$ is dumped, and $S$ is 
refined with the implementation of $v$ before the next circuit is computed.  The 
final $f_v$ are in CNF, so the circuits have a depth of only two.  Even after 
optimizations and mapping to standard cells, the depth usually remains 
low~\cite{EhlersKH12}, which enables fast clock rates.

Once $\neg S$ is available in CNF, the algorithm only adds clauses to existing 
CNFs (i.e., to $R$ and $f_v$).  Only for the resubstitution in 
line~\ref{alg:qbfl:resub}, a CNF encoding of $\neg f_v$ is needed.

\textbf{Optimizations for Safety Specifications.}  As in 
Sect.~\ref{sec:qbfcert}, $\neg S$ is defined as $W \wedge T \wedge \neg W'$. 
This requires a CNF encoding of $\neg W'$.  While computing $\neg W'$ with 
\textsc{NegLearn} is beneficial for \qbfcert, it does not pay off for 
\textsc{SyLearnQBF}. Hence, we build a CNF for $\neg W'$ with one auxiliary 
variable per clause of $W'$.  Recently, the QBF solver \depqbf was equipped 
with incremental solving capabilities~\cite{LonsingE14}. \textsc{SyLearnQBF} 
is well suited for incremental solving. We use two solver instances for 
line~\ref{alg:qbfl:check} and~\ref{alg:qbfl:core} respectively.  For each $v\in 
\overline{v}$, a new incremental session is started.  During the inner loop, we 
only add clauses to the former solver. The QBF of the latter even stays the 
same.  \depqbf supports unsatisfiable cores natively.  The resulting cores are 
small but not necessarily minimal, so we iterate over the remaining literals to 
obtain even smaller cores because (slightly) smaller cores typically mean (much) 
less iterations.

\subsection{Interpolation} \label{sec:interpol}

Jiang et al.~\cite{JiangLH09} present two interpolation-based approaches to 
synthesize circuits for one $v\in \overline{v}$ after the other.  The first one 
expands $S$ over $\overline{v}$.  We consider this intractable in our setting. 
The second approach circumvents the quantifier alternation and expansion by 
temporarily treating output signals as inputs:
\begin{algorithmic}[1]
\Procedure{SyInt}{$S(\overline{x},\overline{i},\overline{o},
                        \overline{x}')$}
  \State $\overline{d} := \overline{x} \cup \overline{i} \cup
                          \overline{o} \cup \overline{x}'$,\;\;
         $\overline{r} := \emptyset$
  \For{$v\in \overline{v}$}
    \State $\overline{d} := \overline{d} \setminus \{v\}$,\;
           $\overline{r} := \overline{r} \cup \{v\}$
    \State $\overline{r}_1,\overline{r}_2,\overline{r}_3,\overline{r}_4 :=
           \textsf{create4FreshCopies}(\overline{r})$
    \State $M_1(\overline{d},\overline{r}_1,\overline{r}_2) :=
           (S\wedge v)[\overline{r}\subby\overline{r}_1]
           \wedge
           (\neg S \wedge \neg v)[\overline{r}\subby\overline{r}_2]$
    \State $M_0(\overline{d},\overline{r}_3,\overline{r}_4) :=
           (S\wedge \neg v)[\overline{r}\subby\overline{r}_3]
           \wedge 
           (\neg S \wedge v)[\overline{r}\subby \overline{r}_4]$
    \State $f_v(\overline{d}) := \interpol(
              M_1(\overline{d},\overline{r}_1,\overline{r}_2),
              M_0(\overline{d},\overline{r}_3,\overline{r}_4)
           )$\label{alg:int:int}
    \State $\textsc{dumpCircuit}(v, f_v)$,\;\;
           $S := S \wedge (v\leftrightarrow f_v)$ \label{alg:int:resub}
  \EndFor
\EndProcedure
\end{algorithmic}
In each iteration, $\overline{d}$ contains all variables on which the 
implementation of the current $v\in \overline{v}$ can depend, and $\overline{r}$ 
contains the rest.  For $\overline{v}=(v_1,\ldots,v_n)$, $v_1$ can 
depend not only on $\overline{u}$ but also on $(v_2,\ldots,v_n)$, $v_2$ can 
depend on $\overline{u}$ and $(v_3,\ldots,v_n)$, etc.  Yet, when the circuits 
for all $v\in \overline{v}$ are built together, the signals $\overline{v}$ 
effectively depend on $\overline{u}$ only. The formulas $M_1$ and $M_0$ 
characterize the $\overline{d}$-vectors for which $v$ must be set to
$\true$ and $\false$ respectively.  The syntax 
$[\overline{r}\subby\overline{r}_i]$ means that the variables $\overline{r}$ are 
renamed by fresh copies $\overline{r}_i$. Line~\ref{alg:int:int} computes an 
interpolant between $M_1$ and $M_0$.  The property $M_1 \Rightarrow f_v 
\Rightarrow \neg M_0$ of the interpolant ensures that (a) $f_v$ is $\true$ 
whenever it must be $\true$, and (b) whenever $f_v$ is $\true$ then it does not 
have to be $\false$.  The renaming of the variables $\overline{r}$ has the 
effect that $f_v$ can only depend on the shared signals~$\overline{d}$.

\textbf{Optimizations for Safety Specifications.} In order to avoid
double-negations of $W$ in $S$ by negating $S$, we compute
\begin{gather*}
M_1 := (T \wedge W' \wedge v)[\overline{r}\subby\overline{r}_1] \wedge
         (T \wedge \neg v \wedge W \wedge \neg W')
         [\overline{r}\subby\overline{r}_2]\\
M_0 := (T \wedge W' \wedge \neg v)[\overline{r}\subby\overline{r}_3] \wedge
         (T \wedge v \wedge W \wedge \neg W')
         [\overline{r}\subby\overline{r}_4]
\end{gather*}
Note the difference to a plain substitution of $S = T \wedge (\neg W \vee W')$ 
and $\neg S = T \wedge W \wedge \neg W'$ in \textsc{SyInt}: $(\neg W \vee W')$ 
reduces to $W'$ due to the conjunction with $W$ from $\neg S$.  This is 
fortunate because disjunctions are expensive in CNF. Since \textsc{SyInt} allows 
$v_i$ to depend on other $v_j$ with $j>i$, it is sensitive to the variable 
order, both regarding execution time and circuit size. We exploit this insight 
with the following optimization. Once $v_i$ has been synthesized, we analyze on 
which $v_j$ it actually depends. If $v_i$ does not depend on a particular $v_j$, 
then $v_j$ is allowed to depend on $v_i$. This gives an increased flexibility 
without introducing circular dependencies.  We simplify all computed 
interpolants with \Abc\footnote{We use the command sequence \texttt{strash; 
refactor -zl; rewrite -zl;} up to 3 times, followed by \texttt{dfraig; rewrite 
-zl; dfraig;}.}~\cite{BraytonM10}.

\subsection{SAT-Based Learning} \label{sec:satlearn}

Here, we use \textsc{SyInt} but with a special interpolation procedure (called 
in line~\ref{alg:int:int}) that applies computational learning:
\begin{algorithmic}[1]
\Procedure{IntLearn}{$M_1(\overline{d},\overline{r}_1,\overline{r}_2)$,
                     $M_0(\overline{d},\overline{r}_3,\overline{r}_4)$}
  \State $f := \true$
    \While{$\mathsf{sat}$, with
      $(\mathsf{sat},\mathbf{d}):=\propsatmodel(M_0 \wedge f)$}
          \label{alg:satl:check}
      \State $f := f \wedge \neg \propsatcore(\mathbf{d}, M_1)$
      \label{alg:satl:core}
    \EndWhile
    \State \textbf{return} $f$
\EndProcedure
\end{algorithmic}
As long as there exists some $\mathbf{d}$ for which $f$ is $\true$ but must be 
$\false$, i.e., $M_0 \wedge f$ is satisfiable, we refine $f$ with an additional 
clause that excludes the cube $\mathbf{d}$ witnessing this insufficiency.  By 
taking the unsatisfiable core, other inputs are also mapped to $\false$ as long 
as $\false$ is allowed.

\textbf{Optimizations.}
We use two SAT solver instances, one for line~\ref{alg:satl:check} and one for 
line~\ref{alg:satl:core}.  A new incremental session is started upon each call 
of \textsc{IntLearn}.  Using activation variables to set 
$\overline{v}$-variables to $\true$, $\false$, or equal to their renamed copy, 
we can even use one incremental session throughout the entire \textsc{SyInt} 
procedure. However, this did not result in significant improvements in our 
experiments.  All optimizations discussed in Sect.~\ref{sec:interpol} can be 
applied.  We also extended the variable dependency optimization further:  The 
CNF $T$ often contains many auxiliary variables that are defined uniquely by 
other signals of $\overline{x}$, $\overline{i}$, $\overline{o}$.  If some of 
these auxiliary variables depend only on $\overline{d}$, then we allow $f$ to 
depend on them as well by including them into $\overline{d}$.  This can be 
beneficial because these auxiliary variables often capture the important 
connections between the variables $\overline{x}$, $\overline{i}$, 
$\overline{o}$.  When dumping the circuits, we add additional gates that define 
the referenced auxiliary variables as done by $T$.  We also implemented a second 
minimization pass that tries to remove every clause and literal from every CNF 
$f$ after \textsc{SyInt} is done.  However, this only results in minor circuit 
size improvements (around 20\%).

\section{Experimental Results} \label{sec:exp}

\subsection{Implementation}
We implemented the discussed methods and optimizations in the SAT-based 
synthesis tool \demiurge%
\footnote{\url{
http://www.iaik.tugraz.at/content/research/design_verification/demiurge/}.}
\cite{BloemKS14}.  \demiurge synthesizes 
\aiger\footnote{\url{http://fmv.jku.at/aiger/}} circuits from safety 
specifications and complies with the 
\textsf{SyntComp}\footnote{\url{http://www.syntcomp.org/}} competition rules. 
The archive of version 1.1.0 contains way more experiments than reported here. 
E.g., for the SAT-based learning approach alone we implemented 24 variants.  
Here, we only compare interesting versions, summarized in the following table.
\begin{center}
\begin{tabular}{c|c|l}
Name & Engine       & Algorithm\\
\hline
\textsf{BDD}&\cudd  2.4.2   & Cofactor-Based~\cite{BloemGJPPW07}\\
\textsf{QC} &\qbfcert 1.0   & QBF-Certification (Sect~\ref{sec:qbfcert})\\
\textsf{QL} &\depqbf  3.02  & \textsc{SyLearnQBF} (Sect~\ref{sec:qbflearn})\\
\textsf{SI} &\mathsat 5     & \textsc{SyInt} (Sect~\ref{sec:interpol})\\
\textsf{SL} &\lingeling ats & \textsc{SyInt}+\textsc{IntLearn}
                               (Sect~\ref{sec:satlearn})\\
\textsf{SLN}&\lingeling ats & SL without dependency opt.
\end{tabular}
\end{center}
\textsf{BDD} serves as baseline for our comparison. It was created by students 
and won a competition held in a synthesis lecture. It implements a 
cofactor-based approach~\cite{BloemGJPPW07}, uses dynamic variable reordering, 
and forced reorderings at certain points. \textsf{QC}, \textsf{QL}, \textsf{SI}, 
and \textsf{SL} implement the methods from the previous section with all 
optimizations.  \textsf{SLN} is used to highlight the benefits of the dependency 
optimization.  All our methods use \Abc\footnotemark[5]~\cite{BraytonM10} to 
minimize the final circuits further.  \textsf{SI} uses \mathsat, which supports 
several interpolation schemes. We use McMillan's scheme 
(see~\cite{DSilvaKPW10}), but the performance is similar with other schemes.  We 
also implemented our own interpolation engine by processing proofs produced by 
\picosat.  However, the proof files grew prohibitively large.

The limitations of our implementation are that it can only handle safety 
specifications in \aiger format, it can produce circuit only in \aiger format, 
and it runs under Linux only.

\subsection{Benchmarks}

We use the same benchmarks as~\cite{BloemKS14}, but report here only results for 
the interesting ones.  The benchmarks \texttt{amba}$ij$ specify an arbiter 
for ARM's AMBA AHB bus~\cite{BloemGJPPW07}, where $i$ is the number of masters, 
and $j\in\{\texttt{c},\texttt{b},\texttt{f}\}$ indicates the method used to 
transform the original benchmark~\cite{BloemGJPPW07} into our input 
format~\cite{BloemKS14}.  The benchmarks \texttt{genbuf}$ij$, again with 
$j\in\{\texttt{c},\texttt{b},\texttt{f}\}$, define a generalized 
buffer~\cite{BloemGJPPW07} connecting $i$ senders to two receivers.  The 
specifications \texttt{add}$i$ and \texttt{mult}$i$ denote $i$-bit combinational 
adders and multipliers.

\subsection{Results and Discussion}

\begin{figure}[htb]
\centering
   \subfigure[Execution time for \texttt{amba} and
              \texttt{genbuf}.\label{fig:cactus_amba_gen}]
             {\includegraphics[width=0.48\textwidth]{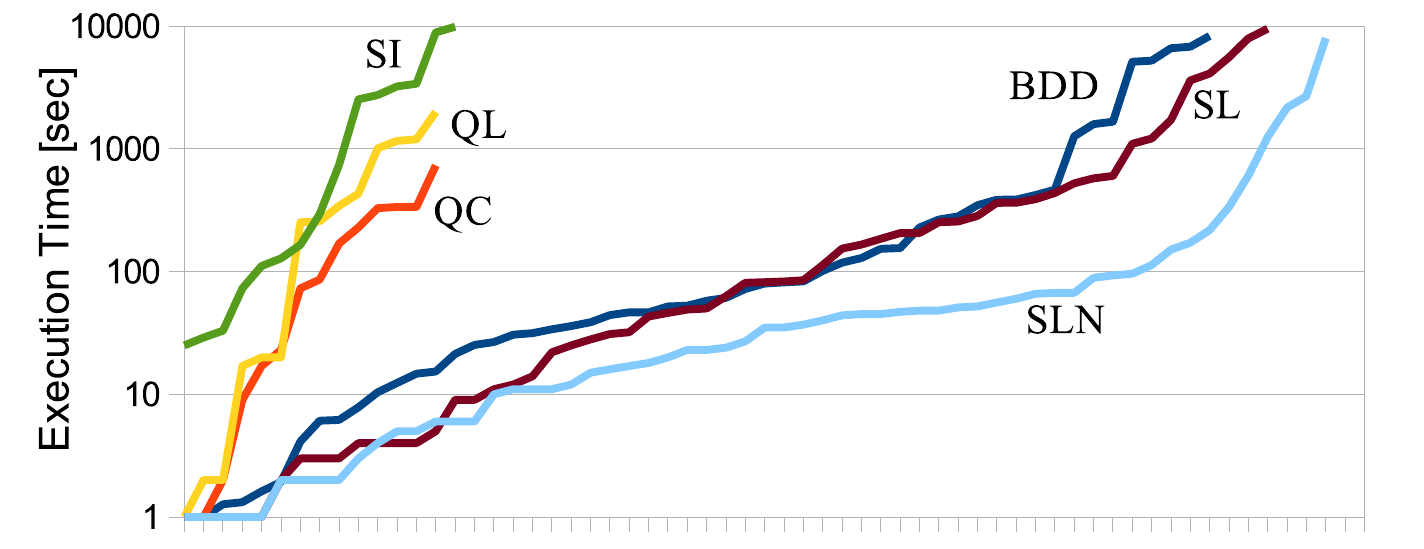}}
   \subfigure[Execution time for \texttt{add} and
              \texttt{mult}.\label{fig:cactus_add_mult}]
             {\includegraphics[width=0.48\textwidth]{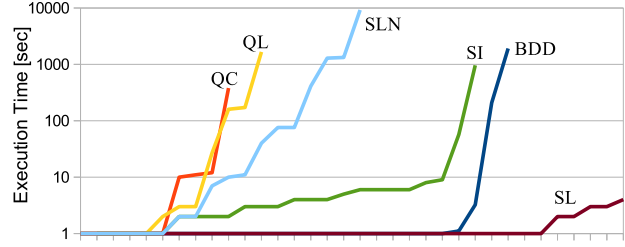}}
   \subfigure[Circuit size for all benchmarks.\label{fig:cactus_size}]
             {\includegraphics[width=0.48\textwidth]{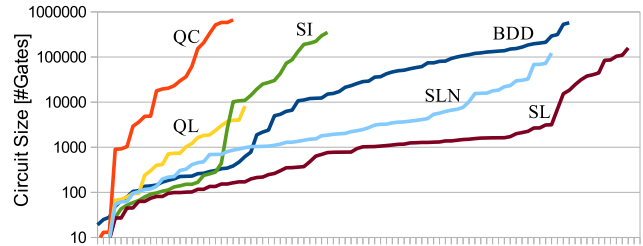}}
   \caption{Cactus plots summarizing our performance evaluation.}
  \label{fig:cactus}
\end{figure}

Fig.~\ref{fig:cactus} summarizes our results with cactus plots. The y-axis gives 
the execution time or circuit size, and the x-axis gives the number of 
benchmarks that can be solved within this time or size limit. Concrete numbers 
and more plots can be found in the appendix and in the downloadable archive.  
All experiments were performed on an Intel 
Xeon E5430 CPU running a 64 bit Linux at $2.66$\,GHz. The memory limit was set 
to $8$\,GB, the time-out to $10\,000$ seconds. All circuits have been 
successfully model checked.

Method \textsf{SL} achieves the best results both regarding execution time and 
circuit size.  The dependency optimization (\textsf{SL} vs.~\textsf{SLN}) is 
very beneficial for \texttt{add} and \texttt{mult}, but slower for \texttt{amba} 
and \texttt{genbuf}.  \textsf{QC}, \textsf{QL}, and \textsf{SI} do not perform 
so well.  Using incremental QBF solving in \textsf{QL} gives an average speedup 
of factor $3.5$.  The speedup factor compared to using the QBF preprocessor 
\bloqqer is even $21$.  Still, \textsf{QL} is not very competitive. \textsf{BDD} 
is much better, but still outperformed by \textsf{SL}. In particular, 
\textsf{SL} outperforms \textsf{SI} by many orders of magnitude. Hence, our idea 
of implementing the interpolation procedure with computational learning is very 
beneficial.  Execution time and circuit size are not in conflict but rather 
correlate.  The time for optimization with \Abc is usually insignificant, but 
only yields moderate size reductions (around $25\,\%$ for SL).
Using method \textsf{SLN}, \demiurge won a track of 
\textsf{SyntComp} 2014.
One reason was the small circuit size compared to other tools.

\section{Conclusion} \label{sec:concl}

We compared several SAT- and QBF-based methods to synthesize circuits from 
strategies, and presented optimizations and efficient implementations for safety 
specifications.  Our SAT-based learning method combines the quantifier 
elimination approach by Jiang et al.~\cite{JiangLH09} with computational 
learning as proposed by Ehlers et al.~\cite{EhlersKH12}, and outperforms BDDs 
both regarding execution time and circuit size in our experiments.

Future research includes preprocessing for incremental QBF solving, exploiting 
unreachable states, and parallelization.


\bibliography{references}
\begin{table*}
\appendix
Table~\ref{table:appendix} contains more performance results.
``T'' indicates a time-out of $10\,000$ seconds. The suffix $k$ stands for
a multiplication with $1000$.  The size column $G$ gives the number of
\aiger gates defining $T$.  The column ``files'' in \textsf{QC} gives the
size of the intermediate files (the QBF trace) produced by \qbfcert;  we
aborted at $20$\,GB. ``M'' indicates that \Abc ran out of
memory (because \textsf{QC} produces huge circuits).  More details can be
found in the downloadable archive.
\end{table*}

\begin{table*}
\setlength{\tabcolsep}{4.7pt}
\renewcommand{\arraystretch}{0.95}
\begin{center}
\caption{Performance results.}
\label{table:appendix}
{\scriptsize
\begin{tabular}{lccccccccccccccccccccccccccccccc}
\toprule[1.3pt]
                     &\multicolumn{4}{c}{Size}
                     &&\multicolumn{2}{c}{\textsc{BDD}}
                     &&\multicolumn{3}{c}{QC}
                     &&\multicolumn{2}{c}{QL}
                     &&\multicolumn{2}{c}{SI}
                     &&\multicolumn{2}{c}{SL}
                     &&\multicolumn{2}{c}{SLN}
                     \\
\cmidrule{2-5}
\cmidrule{7-8}
\cmidrule{10-12}
\cmidrule{14-15}
\cmidrule{17-18}
\cmidrule{20-21}
\cmidrule{23-24}
                     &$|\overline{i}|$
                     &$|\overline{c}|$
                     &$|\overline{x}|$
                     &$G$
                     &
                     &time
                     &size
                     &
                     &time
                     &size
                     &files
                     &
                     &time
                     &size
                     &
                     &time
                     &size
                     &
                     &time
                     &size
                     &
                     &time
                     &size
                     \\
\cmidrule{1-5}
\cmidrule{7-8}
\cmidrule{10-12}
\cmidrule{14-15}
\cmidrule{17-18}
\cmidrule{20-21}
\cmidrule{23-24}
                    &[-]    &[-]  &[-]    &[-]
                    &&[sec] &[cells]
                    &&[sec] &[cells] &[MB]
                    &&[sec] &[cells]
                    &&[sec] &[cells]
                    &&[sec] &[cells]
                    &&[sec] &[cells]
\\
\cmidrule{1-5}
\cmidrule{7-8}
\cmidrule{10-12}
\cmidrule{14-15}
\cmidrule{17-18}
\cmidrule{20-21}
\cmidrule{23-24}
 \texttt{add2}      &4 &2 &2 &17  &&0.1  &2    &&1    &13   &0.1    &&0.1  &9    &&1    &9    &&1    &9    &&0.1  & 9    \\
 \texttt{add4}      &8 &4 &2 &45  &&0.1  &4    &&1    &919  &0.3    &&0.1  &94   &&1    &28   &&1    &27   &&0.1  & 104  \\
 \texttt{add6}      &12&6 &2 &73  &&0.1  &6    &&11   &20k  &11     &&3.0  &739  &&1    &43   &&1    &45   &&2    & 688  \\
 \texttt{add8}      &16&8 &2 &101 &&0.1  &8    &&164  &M    &738    &&172  &4.0k &&2    &59   &&1    &63   &&10   & 3.6k \\
 \texttt{add10}     &20&10&2 &129 &&0.1  &10   &&T    &-    &$>$17k &&T    &-    &&2    &79   &&1    &81   &&76   & 16k  \\
 \texttt{add12}     &24&12&2 &157 &&0.1  &12   &&-    &-    &$>$20k &&T    &-    &&3    &97   &&1    &99   &&1.3k & 69k  \\
 \texttt{add14}     &28&14&2 &185 &&0.1  &14   &&-    &-    &$>$20k &&T    &-    &&4    &113  &&1    &117  &&T    & -    \\
 \texttt{add16}     &32&16&2 &213 &&0.1  &16   &&-    &-    &$>$20k &&T    &-    &&4    &132  &&1    &135  &&T    & -    \\
 \texttt{add18}     &36&18&2 &241 &&0.1  &18   &&-    &-    &$>$20k &&T    &-    &&6    &151  &&1    &153  &&T    & -    \\
 \texttt{add20}     &40&20&2 &269 &&0.2  &20   &&-    &-    &$>$20k &&T    &-    &&6    &167  &&1    &171  &&T    & -    \\
\cmidrule{1-5}
\cmidrule{7-8}
\cmidrule{10-12}
\cmidrule{14-15}
\cmidrule{17-18}
\cmidrule{20-21}
\cmidrule{23-24}
 \texttt{mult2}     &4 &4 &0 &24  &&0.1  &4    &&1    &8    &0.1    &&0.1  &8    &&1    &8    &&1    &8    &&0.1  & 8    \\
 \texttt{mult4}     &8 &8 &0 &128 &&0.1  &8    &&1    &1.0k &6.1    &&2    &414  &&3    &440  &&1    &95   &&1    & 412  \\
 \texttt{mult5}     &10&10&0 &217 &&0.2  &10   &&12   &4.9k &142    &&27   &1.9k &&9    &2.6k &&1    &163  &&7    & 1.8k \\
 \texttt{mult6}     &12&12&0 &322 &&1.1  &12   &&380  &23k  &2.9k   &&1.7k &8.2k &&57   &14k  &&1    &247  &&40   & 7.7k \\
 \texttt{mult7}     &14&14&0 &455 &&3.2  &14   &&-    &-    &$>$20k &&T    &-    &&962  &73k  &&1    &351  &&411  & 30k  \\
 \texttt{mult8}     &16&16&0 &604 &&209  &16   &&-    &-    &$>$20k &&T    &-    &&T    &-    &&1    &477  &&9.2k & 122k \\
 \texttt{mult10}    &20&20&0 &964 &&T    &-    &&-    &-    &$>$20k &&T    &-    &&T    &-    &&1    &777  &&T    & -    \\
 \texttt{mult12}    &24&24&0 &1379&&T    &-    &&-    &-    &$>$20k &&T    &-    &&T    &-    &&2    &1.2k &&T    & -    \\
 \texttt{mult16}    &32&32&0 &2450&&T    &-    &&-    &-    &$>$20k &&T    &-    &&T    &-    &&4    &2.1k &&T    & -    \\
\cmidrule{1-5}
\cmidrule{7-8}
\cmidrule{10-12}
\cmidrule{14-15}
\cmidrule{17-18}
\cmidrule{20-21}
\cmidrule{23-24}
 \texttt{genbuf1c}  &5 &6 &21&134 &&0.5  &1.7k &&1    &2.9k &2.4    &&2    &95   &&29   &10k  &&1    &101  &&1    & 62   \\
 \texttt{genbuf2c}  &6 &7 &24&169 &&1.6  &4.8k &&9    &18k  &24.3   &&20   &239  &&73   &25k  &&3    &198  &&2    & 119  \\
 \texttt{genbuf3c}  &7 &9 &27&202 &&6.1  &11k  &&86   &62k  &166    &&257  &709  &&3.2k &191k &&4    &301  &&3    & 224  \\
 \texttt{genbuf4c}  &8 &10&30&242 &&12   &15k  &&329  &208k &1.3k   &&2.0k &1.0k &&9.9k &222k &&14   &362  &&4    & 326  \\
 \texttt{genbuf5c}  &9 &12&33&284 &&21   &21k  &&1.4k &M    &4.7k   &&T    &-    &&T    &-    &&43   &779  &&11   & 455  \\
 \texttt{genbuf6c}  &10&13&35&323 &&36   &36k  &&-    &-    &$>$20k &&T    &-    &&T    &-    &&63   &781  &&17   & 694  \\
 \texttt{genbuf8c}  &12&15&40&406 &&282  &61k  &&-    &-    &$>$20k &&T    &-    &&T    &-    &&437  &1.0k &&44   & 1.0k \\
 \texttt{genbuf10c} &14&18&45&494 &&155  &100k &&-    &-    &$>$20k &&T    &-    &&T    &-    &&206  &1.6k &&45   & 1.5k \\
 \texttt{genbuf12c} &16&20&49&561 &&266  &154k &&-    &-    &$>$20k &&T    &-    &&T    &-    &&166  &1.4k &&51   & 1.1k \\
 \texttt{genbuf16c} &20&24&58&733 &&465  &203k &&-    &-    &$>$20k &&T    &-    &&T    &-    &&389  &1.5k &&47   & 2.0k \\
\cmidrule{1-5}
\cmidrule{7-8}
\cmidrule{10-12}
\cmidrule{14-15}
\cmidrule{17-18}
\cmidrule{20-21}
\cmidrule{23-24}
 \texttt{genbuf1b}  &5 &6 &23&141 &&0.8  &2.3k &&1    &3.6k &3.1    &&1    &70   &&25   &11k  &&1    &102  &&1    & 62   \\
 \texttt{genbuf2b}  &6 &7 &26&174 &&6.2  &6.5k &&17   &30k  &31.7   &&17   &399  &&111  &19k  &&1    &213  &&1    & 116  \\
 \texttt{genbuf3b}  &7 &9 &30&208 &&7.8  &11k  &&73   &154k &354    &&252  &1.6k &&2.5k &202k &&3    &262  &&2    & 217  \\
 \texttt{genbuf4b}  &8 &10&33&245 &&34   &26k  &&733  &666k &2.3k   &&1.2k &1.2k &&3.4k &352k &&2    &353  &&2    & 307  \\
 \texttt{genbuf5b}  &9 &12&37&282 &&52   &29k  &&T    &-    &16k    &&T    &-    &&T    &-    &&5    &693  &&5    & 863  \\
 \texttt{genbuf6b}  &10&13&40&322 &&27   &23k  &&-    &-    &$>$20k &&T    &-    &&T    &-    &&11   &1.3k &&11   & 2.2k \\
 \texttt{genbuf8b}  &12&15&46&395 &&61   &51k  &&-    &-    &$>$20k &&T    &-    &&T    &-    &&12   &1.1k &&6    & 804  \\
 \texttt{genbuf10b} &14&18&53&475 &&118  &80k  &&-    &-    &$>$20k &&T    &-    &&T    &-    &&25   &775  &&16   & 1.9k \\
 \texttt{genbuf12b} &16&20&59&547 &&31   &74k  &&-    &-    &$>$20k &&T    &-    &&T    &-    &&46   &1.0k &&27   & 2.0k \\
 \texttt{genbuf16b} &20&24&71&687 &&44   &107k &&-    &-    &$>$20k &&T    &-    &&T    &-    &&113  &1.3k &&113  & 18k  \\
\cmidrule{1-5}
\cmidrule{7-8}
\cmidrule{10-12}
\cmidrule{14-15}
\cmidrule{17-18}
\cmidrule{20-21}
\cmidrule{23-24}
 \texttt{genbuf1f}  &5 &6 &23&138 &&1.3  &2.0k &&2    &4.8k &2.9    &&2    &67   &&33   &11k  &&1    &74   &&0.1  & 54   \\
 \texttt{genbuf2f}  &6 &7 &26&168 &&1.9  &6.1k &&23   &36k  &33     &&20   &301  &&128  &27k  &&1    &218  &&0.1  & 136  \\
 \texttt{genbuf3f}  &7 &9 &30&200 &&82   &12k  &&169  &332k &673    &&432  &1.8k &&2.7k &133k &&4    &372  &&2    & 198  \\
 \texttt{genbuf4f}  &8 &10&33&235 &&15   &17k  &&2.7k &M    &6.0k   &&T    &-    &&T    &-    &&9    &762  &&10   & 476  \\
 \texttt{genbuf5f}  &9 &12&37&272 &&47   &46k  &&-    &-    &$>$20k &&T    &-    &&T    &-    &&49   &1.6k &&18   & 1.0k \\
 \texttt{genbuf6f}  &10&13&40&309 &&53   &42k  &&-    &-    &$>$20k &&T    &-    &&T    &-    &&185  &2.7k &&67   & 1.9k \\
 \texttt{genbuf7f}  &11&14&43&344 &&386  &112k &&-    &-    &$>$20k &&T    &-    &&T    &-    &&524  &3.1k &&172  & 2.5k \\
\cmidrule{1-5}
\cmidrule{7-8}
\cmidrule{10-12}
\cmidrule{14-15}
\cmidrule{17-18}
\cmidrule{20-21}
\cmidrule{23-24}
 \texttt{amba2c}    &7 &8 &28&177 &&1.3  &5.2k &&229  &516k &985    &&343  &2.3k &&294  &43k  &&4    &1.2k &&6    & 1.2k \\
 \texttt{amba3c}    &9 &10&34&237 &&10   &15k  &&-    &-    &$>$20k &&T    &-    &&8.9k &296k &&22   &2.3k &&12   & 1.3k \\
 \texttt{amba4c}    &11&11&38&279 &&229  &129k &&-    &-    &$>$20k &&T    &-    &&T    &-    &&206  &18k  &&93   & 10k  \\
 \texttt{amba5c}    &13&13&43&345 &&1.7k &166k &&-    &-    &$>$20k &&T    &-    &&T    &-    &&256  &15k  &&35   & 3.2k \\
 \texttt{amba6c}    &15&14&47&395 &&5.1k &131k &&-    &-    &$>$20k &&T    &-    &&T    &-    &&576  &24k  &&40   & 3.8k \\
 \texttt{amba7c}    &17&15&52&449 &&6.8k &138k &&-    &-    &$>$20k &&T    &-    &&T    &-    &&1.2k &38k  &&67   & 4.3k \\
 \texttt{amba9c}    &21&18&61&583 &&T    & -   &&-    &-    &$>$20k &&T    &-    &&T    &-    &&4.1k &86k  &&151  & 5.4k \\
 \texttt{amba10c}   &23&19&65&630 &&1.6k &212k &&-    &-    &$>$20k &&T    &-    &&T    &-    &&5.6k &110k &&218  & 7.0k \\
\cmidrule{1-5}
\cmidrule{7-8}
\cmidrule{10-12}
\cmidrule{14-15}
\cmidrule{17-18}
\cmidrule{20-21}
\cmidrule{23-24}
 \texttt{amba2b}    &7 &8 &31&189 &&15   &12k  &&337  &584k &1.3k   &&1.2k &3.7k &&739  &87k  &&4    &1.3k &&6    & 1.1k \\
 \texttt{amba3b}    &9 &10&36&231 &&83   &74k  &&-    &-    &$>$20k &&T    &-    &&T    &-    &&28   &3.1k &&11   & 1.4k \\
 \texttt{amba4b}    &11&11&42&286 &&8.3k &314k &&-    &-    &$>$20k &&T    &-    &&T    &-    &&602  &31k  &&336  & 18k  \\
 \texttt{amba5b}    &13&13&47&344 &&6.6k &574k &&-    &-    &$>$20k &&T    &-    &&T    &-    &&1.1k &41k  &&45   & 4.1k \\
 \texttt{amba6b}    &15&14&52&391 &&T    &-    &&-    &-    &$>$20k &&T    &-    &&T    &-    &&3.6k &85k  &&60   & 5.7k \\
 \texttt{amba7b}    &17&15&57&438 &&T    &-    &&-    &-    &$>$20k &&T    &-    &&T    &-    &&8.0k &159k &&96   & 6.6k \\
 \texttt{amba9b}    &21&18&68&558 &&T    &-    &&-    &-    &$>$20k &&T    &-    &&T    &-    &&T    &-    &&605  & 22k  \\
 \texttt{amba10b}   &23&19&73&606 &&T    &-    &&-    &-    &$>$20k &&T    &-    &&T    &-    &&T    &-    &&2.7k & 72k  \\
\cmidrule{1-5}
\cmidrule{7-8}
\cmidrule{10-12}
\cmidrule{14-15}
\cmidrule{17-18}
\cmidrule{20-21}
\cmidrule{23-24}
 \texttt{amba2f}   &7 &8 &31&181 &&4.1  &12k  &&336  &584k &1.4k   &&1.0k &3.0k &&165  &30k  &&3    &1.3k &&5    & 1.0k \\
 \texttt{amba3f}   &9 &10&37&229 &&101  &36k  &&T    &-    &$>$16k &&T    &-    &&T    &-    &&83   &7.0k &&52   & 6.2k \\
 \texttt{amba4f}   &11&11&43&282 &&1.3k &293k &&-    &-    &$>$20k &&T    &-    &&T    &-    &&1.7k &44k  &&1.3k & 33k  \\
 \texttt{amba5f}   &13&13&49&346 &&T    &-    &&-    &-    &$>$20k &&T    &-    &&T    &-    &&9.5k &104k &&2.2k & 23k  \\
 \texttt{amba6f}   &15&14&54&391 &&T    &-    &&-    &-    &$>$20k &&T    &-    &&T    &-    &&T    &-    &&8.0k & 31k  \\
\bottomrule[1.3pt]   
\end{tabular}        
}
\end{center}
\end{table*}



\begin{figure*}
\centering
   \subfigure[Execution time for \texttt{add}.\label{fig:time_add}]
             {\includegraphics[width=0.48\textwidth]{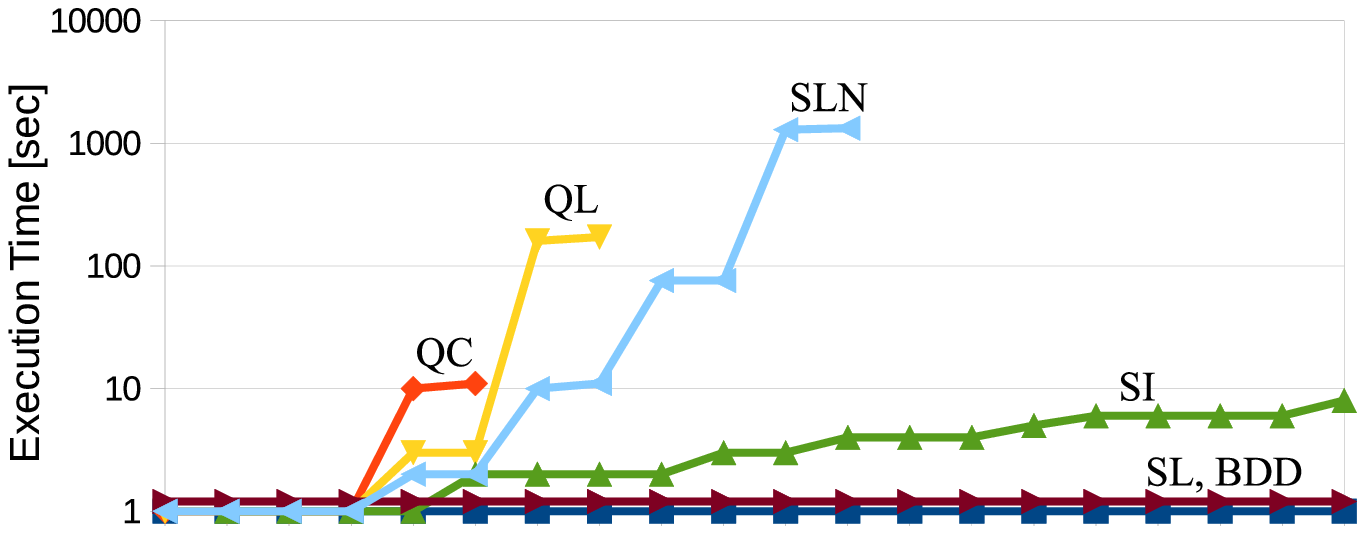}}
   \subfigure[Circuit size for \texttt{add}.\label{fig:size_add}]
             {\includegraphics[width=0.48\textwidth]{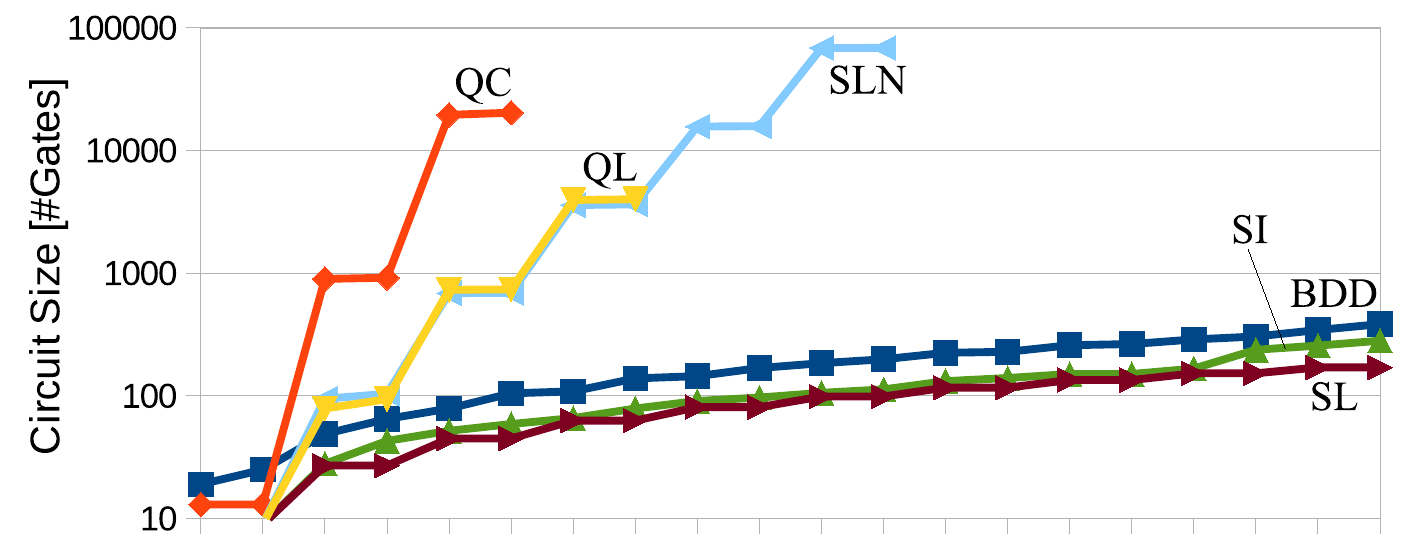}}  
   \subfigure[Execution time for \texttt{mult}.\label{fig:time_mult}]
             {\includegraphics[width=0.48\textwidth]{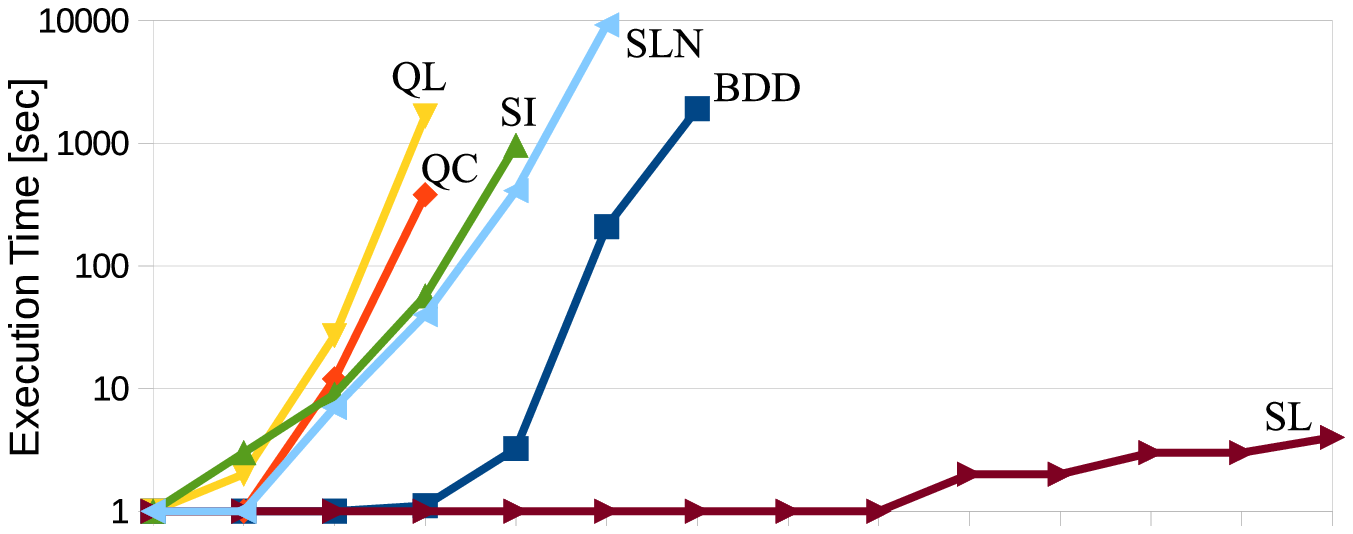}}          
   \subfigure[Circuit size for \texttt{mult}.\label{fig:size_mult}]
             {\includegraphics[width=0.48\textwidth]{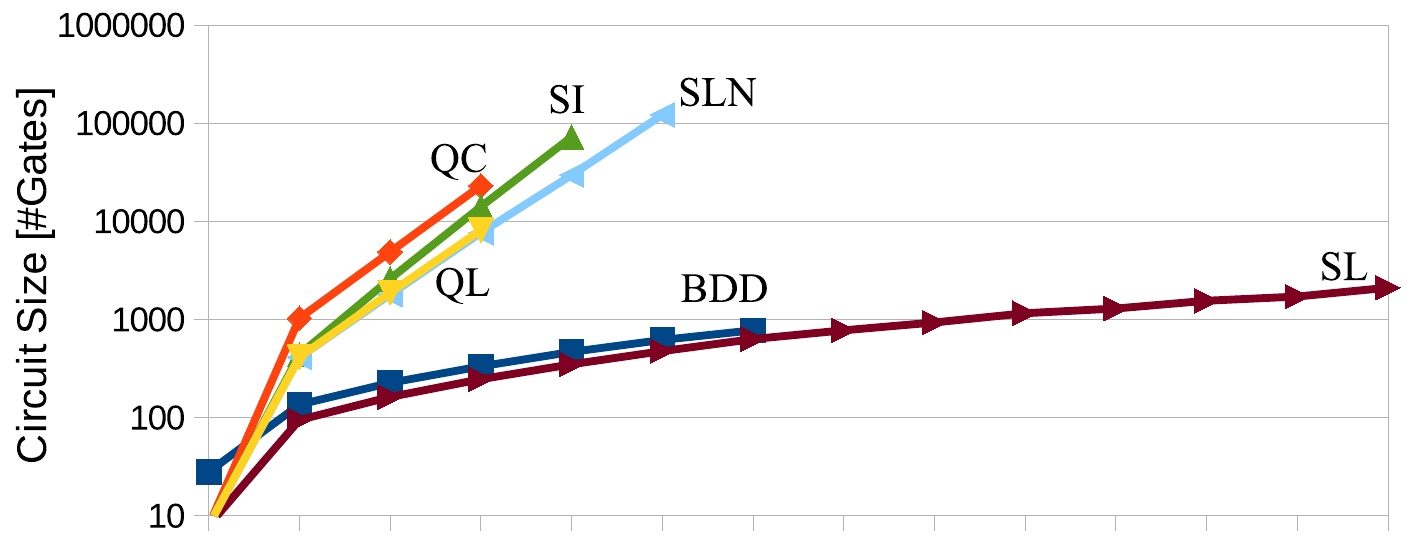}}
   \subfigure[Execution time for \texttt{genbuf}.\label{fig:time_genbuf}]
             {\includegraphics[width=0.48\textwidth]{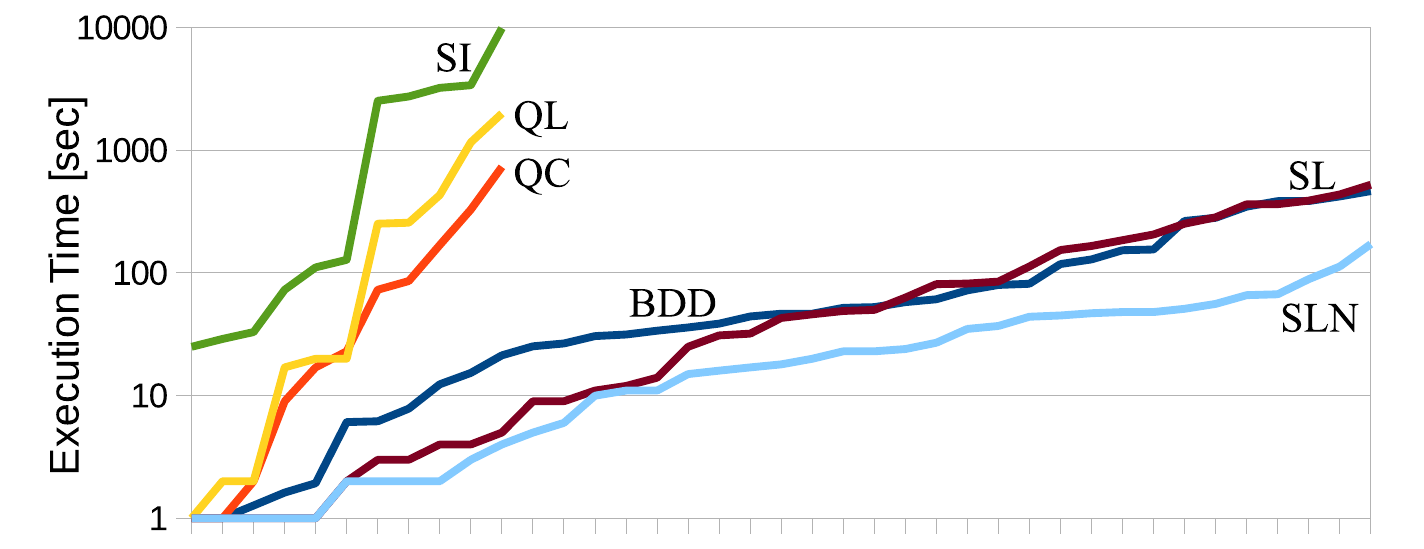}}       
   \subfigure[Circuit size for \texttt{genbuf}.\label{fig:size_genbuf}]
             {\includegraphics[width=0.48\textwidth]{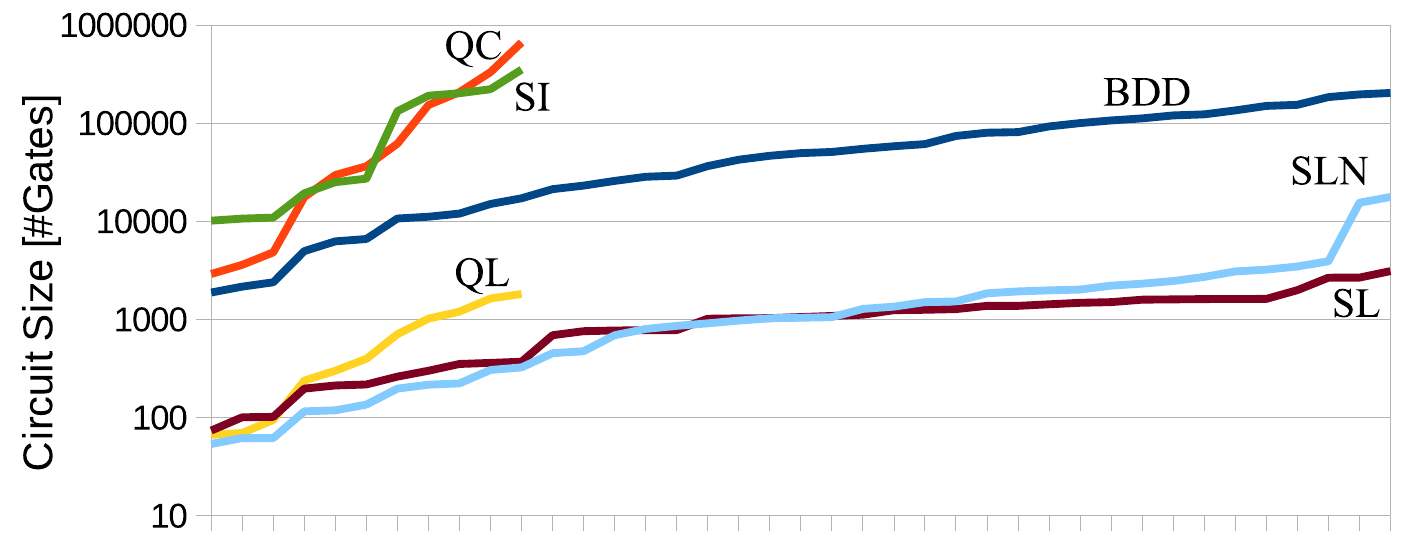}}  
   \subfigure[Execution time for \texttt{amba}.\label{fig:time_amba}]
             {\includegraphics[width=0.48\textwidth]{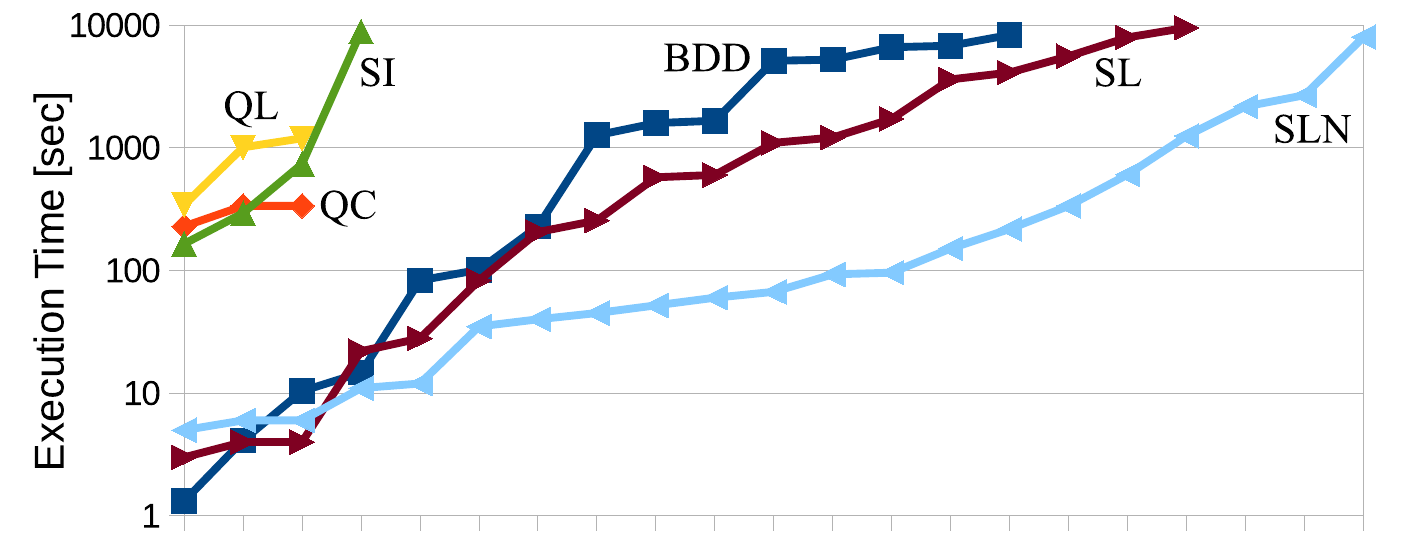}}          
   \subfigure[Circuit size for \texttt{amba}.\label{fig:size_amba}]
             {\includegraphics[width=0.48\textwidth]{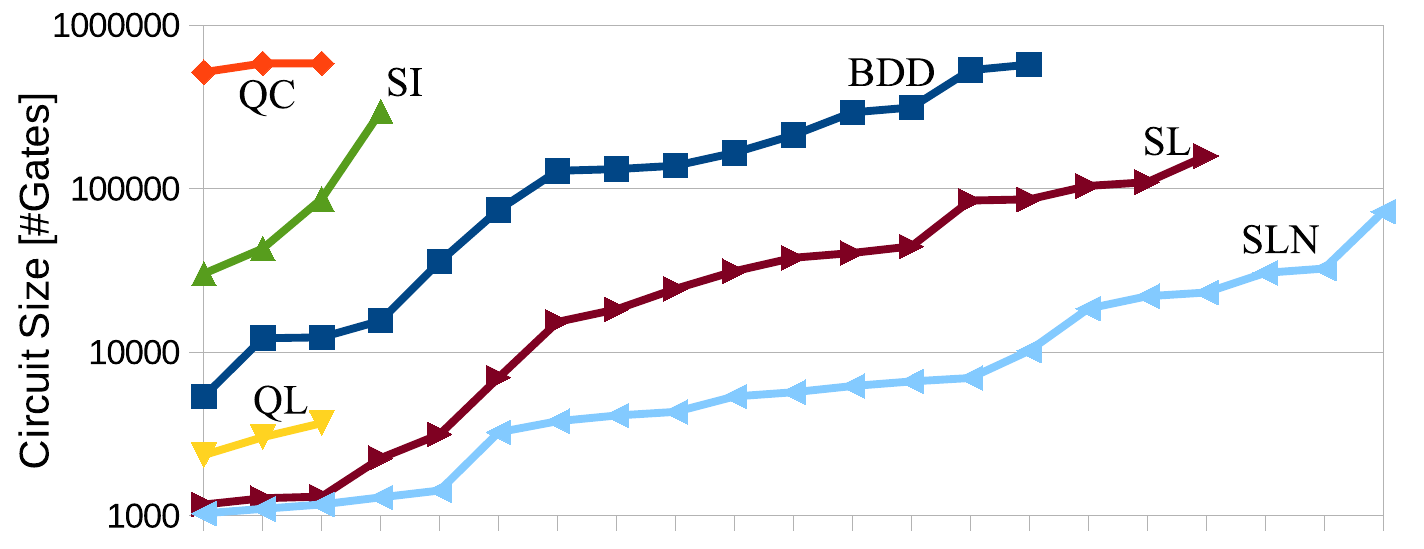}}   
   \caption{Cactus plots summarizing execution time and circuit size per 
            benchmark.}
  \label{fig:cactus_detailed}
\end{figure*}

\noindent
Fig.~\ref{fig:cactus_detailed} contains cactus plots summarizing the execution 
time and circuit size per benchmark.  For \texttt{mult}, the dependency 
optimization has an even stronger positive impact than for \texttt{add}, both 
regarding execution time and circuit size (\textsf{SL} vs.~\textsf{SLN}).  For 
\texttt{genbuf}, it results in smaller circuits, but is slower. For 
\texttt{amba}, it is not beneficial in either metric.  The interpolation-based 
method \textsf{SI} is competitive for \texttt{add}, but has
troubles with the other benchmarks.  The difference in circuit size can grow 
very large (more than three orders of magnitude).  Our a-posteriori circuit 
minimization using \Abc can only compensate an insignificant fraction of 
this difference.  Investing more effort in post-processing can be expensive, 
especially for large circuits.  We therefore conclude that circuit size is 
best considered during the synthesis process already.  The fact that circuit 
size and execution time correlate is not surprising.  Most of our methods 
compute circuits for one output signal after the other. The strategy is refined 
with the circuit for one signal before continuing with the next one. This is 
necessary in order to prevent uncoordinated choices.  If the computed circuits 
are complicated, then this re-substitution makes the strategy formula for the 
next output complicated, which results in higher computation times.  This 
may also explain why implementing an interpolation procedure with computational 
learning is very beneficial (\textsf{SL} vs.~\textsf{SI}): computational 
learning seems to find smaller circuits, and this also pays off in terms of the 
overall computation time.

\end{document}